\documentstyle[a4wide,12pt,epsf]{article}
\begin{document}
\begin{flushright}
\baselineskip=12pt
{SUSX-TH-02-035}\\
{hep-th/0212ddd}\\
{December  2002}
\end{flushright}
\def\IZ{Z\kern-.5em Z}
\begin{center}
{\LARGE \bf INTERSECTING D5-BRANE MODELS 
WITH MASSIVE VECTOR-LIKE LEPTONS \\}
\vglue 0.35cm
{D. BAILIN$^{\clubsuit}$ \footnote
{D.Bailin@sussex.ac.uk}, G. V. KRANIOTIS$^{\spadesuit}$ \footnote
 {G.Kraniotis@sussex.ac.uk, kraniotis@physik.uni-halle.de} and A. LOVE$^{\diamondsuit}$ \\}
\vglue 0.2cm
	{$\clubsuit$ \it  Centre for Theoretical Physics, University of Sussex\\}
{\it Brighton BN1 9QJ, U.K. \\}
{$\spadesuit$ \it Fachbereich Physik, Martin-Luther-Universit\"{a}t Halle-Wittenberg\\
Friedemann-Bach-Platz 6, D-06099, Halle, Germany\\}
{$\diamondsuit$ \it  Centre for Particle Physics, Royal Holloway, University of London \\}
{\it Egham,  Surrey TW20-0EX, U.K. }
\baselineskip=12pt

\vglue 2.5cm
ABSTRACT
\end{center}
We construct eight-stack intersecting D5-brane models, with an orbifold transverse space, that yield the (non-supersymmetric) standard model 
up to vector-like leptons. The matter includes right-chiral neutrinos and the models have the renormalisable Yukawa couplings to tachyonic Higgs doublets 
needed to generate mass terms for {\it all} matter, including the vector-like leptons.
 The models are constrained by the requirement that twisted tadpoles cancel, 
 that the gauge boson coupled to the weak hypercharge $U(1)_Y$ does not get a string-scale mass via a generalised 
Green-Schwarz mechanism, 
and that there are no surviving, unwanted gauged U(1) symmetries coupled to matter.
 Gauge coupling constant ratios close to those measured are easily obtained for reasonable 
values of the parameters, consistently with having the string scale close to the electroweak scale,
 as required to avoid the hierarchy problem. Unwanted (colour-triplet, 
charged-singlet, and neutral-singlet) scalar tachyons can be removed by a suitable choice of the parameters.

{\rightskip=3pc
\leftskip=3pc
\noindent
\baselineskip=20pt

}

\vfill\eject
\setcounter{page}{1}
\pagestyle{plain}
\baselineskip=14pt

The D-brane world offers an attractive, bottom-up route to getting standard-like models from Type II string theory \cite{UTCA}. 
Open strings that begin and end on a stack of $M$ D-branes generate the gauge bosons of the group $U(M)$ living in the world volume of the D-branes.
 So the 
standard approach is to start with one stack of 3 D-branes, another of 2, and $n$ other stacks each having just 1 D-brane, thereby generating 
the gauge group $U(3) \times U(2) \times U(1)^n$. Fermions in bi-fundamental
 representations of the corresponding gauge groups can arise at the intersections of such stacks \cite{BDL}, but to get $D=4$ {\it chiral} fermions 
  the intersecting branes should sit at a singular point in the space transverse to the branes, an orbifold fixed point, for example. In general,
   such configurations yield a non-supersymmetric spectrum, so to avoid the hierarchy problem the
    string scale associated with such models must be no more than a few TeV. Gravitational interactions occur in the bulk ten-dimensional space, and 
    to ensure that the Planck energy has its observed large value, it is necessary that there are large dimensions transverse to the branes \cite{ADD}. 
The D-branes with which we are concerned wrap the 3-space we inhabit and closed 1-, 2- or 3-cycles of a toroidally 
compactified $T^2, \ T^2 \times T^2$ or $T^2 \times T^2 \times T^2$ space. Thus getting the correct Planck scale effectively means that only D4- and D5-brane models are viable, 
since for D6-branes there is no dimension transverse to all of the intersecting branes.  
In a non-supersymmetric theory the cancellation of the closed-string (twisted) Ramond-Ramond (RR) tadpoles does 
{\it not} ensure the cancellation of the Neveu-Schwarz-Neveu-Schwarz (NSNS) tadpoles.
 There is a resulting instability in the complex structure moduli \cite{BKLO}.
 One  way to stabilise some of the (complex structure) moduli is to use an orbifold, rather than a torus, 
for the space wrapped by the D-branes. If the embedding is supersymmetric, then the instabilities are removed. This has been studied \cite{Cvetic},
 using D6-branes, 
 but it has so far proved difficult to get realistic phenomenology consistent with experimental data from such models.
 Unlike D4-brane models on $T^2 \times T^4/\IZ_N$, D5-brane models on $T^4 \times T^2/\IZ_N$ are necessarily non-supersymmetric 
 in the closed string sector \cite{AFIRU1} and contain closed-string tachyons in the twisted sector. 
 These tachyons may be a source of instability of the background \cite{AFIRU2, CIM}. We have nothing to add to current understanding of this point.

During the past year orientifold models with intersecting D6- and D5-branes
have been constructed that yield 
precisely the fermionic spectrum of the standard model (plus three generations of right-chiral neutrinos) \cite{IMR, CIM}. 
(Other recent work on  intersecting brane models, both supersymmetric 
and non-supersymmetric, and their phenomenological implications may be found in \cite{IBM}.)
The spectrum includes open-string 
 $SU(2)_L$ doublet scalar tachyons that may be regarded as the Higgs doublets that break the electroweak symmetry group, but also, unavoidably,
 open-string colour-triplet and charged singlet tachyons either of which is potentially fatal for the phenomenology. 
The wrapping numbers of the various stacks are constrained by the requirement 
 of RR tadpole cancellation, and this ensures the absence of non-abelian anomalies in the emergent low-energy quantum field theory. 
 A generalised Green-Schwarz mechanism ensures that that the gauge bosons 
  associated with all anomalous $U(1)$s acquire string-scale masses \cite{IRU}, but the gauge bosons of some non-anomalous  $U(1)$s
   can also acquire string-scale masses \cite{IMR}; in all such cases the $U(1)$ group survives as a global symmetry.
 Thus we must also   
  ensure the weak hypercharge group $U(1)_Y$ remains a gauge symmetry by requiring that its gauge boson does {\it not} get such a mass. 
  
 In a recent paper \cite{BKL2} we constructed the first semi-realistic intersecting D4-brane orbifold models that satisfy these constraints.
  We found a unique, one-parameter $(n_2)$ family of six-stack, intersecting D4-brane models having three chiral generations
of matter all of which are coupled to the open-string tachyonic Higgs bosons that generate masses when the electroweak symmetry is spontaneously broken.
Such models all have extra vector-like leptons, as well as open-string charged-singlet scalar tachyons.
 They also have at least one surviving (unwanted) coupled, gauged $U(1)$ symmetry after spontaneous symmetry breaking. 
Relaxing the requirement that all matter has the Higgs couplings necessary to generate masses at renormalisable level
 allows us to construct models without unwanted $U(1)$ gauge symmetries. In this case there are also open-string
  colour-triplet scalar tachyons as well as vector-like $d$ quark matter.  
These models predict ratios of the gauge coupling constants that for some values of a parameter are very close to the measured values. 

Intersecting D5-brane models differ in several ways from D4-brane models, 
and in \cite{BKL3} we constructed eight-stack models, without unwanted, gauged $U(1)$ symmetries, which have additional vector-like 
leptonic, but not quark, matter. Although colour-triplet  and charged-singlet scalar tachyons can arise in these models, the former 
can be removed by choosing the distance between certain parallel 1-cycles to be large, and the latter are 
non-tachyonic for particular choices of the wrapping numbers and ratios of the radii of fundamental 1-cycles. However, although the tree-level
Yukawa couplings of the $n_G=3$ generations of chiral matter to the Higgs doublets allow the generation of mass terms for all of the chiral matter,
including right-chiral neutrinos, this is not true of some of the vector-like matter. The eight-stack models in \cite{BKL3} were all constructed 
with certain restrictions on the numbers of stacks in each class, and the question arises as to whether it is possible to do better
 when these restrictions are relaxed. In the present paper we consider the most general eight-stack models, and show that it is.

We use the same general set-up as in \cite{BKL3}. There is an array of D5-brane stacks, each wrapping  closed 
1-cycles in both 2-tori of $T^2 \times T^2$, and situated
 at a fixed point of the transverse $T^2 /\IZ_3$ orbifold.  
 A stack $a$ is specified by two pairs of wrapping numbers $(n_a,m_a)$  and $(\tilde{n}_a,\tilde{m}_a)$ that 
specify the number of times $a$ wraps the basis 1-cycles  in each  $T^2$. 
When $n_a$ and $m_a$ are coprime 
a single copy of the gauge group $U(N_a)$ occurs;  if $n_a$ and $m_a$ have a highest common factor $f_a$ there are $f_a$ copies of  $U(N_a)$, 
or if $m_a$ (or $n_a$) is zero, then there are $n_a$ (or $m_a$) copies of $U(N_a)$.  The 
same thing happens if  $\tilde{n}_a$ and $\tilde{m}_a$  are not coprime.
The number of intersections of stack $a$ with stack $b$ is given by 
  ${\bf I}_{ab}=I_{ab}\tilde{I}_{ab}$, where $I_{ab} \equiv n_am_b-m_an_b$ and 
  $\tilde{I}_{ab}=\tilde{n}_a\tilde{m}_b-\tilde{m}_a\tilde{n}_b$ are the intersection numbers of the corresponding 1-cycles in each torus $T^2$. 
  The generator $\theta$ of the $Z_3$ point group is embedded in the stack of $N_a$ branes as
 $\gamma _{\theta,a} = \alpha ^{p_a}I_{N_a}$, where $\alpha = e^{2\pi i/3}, \ p_a=0,1,2$. 
 The first two stacks $a=1,2$ defined above, that generate a $U(3) \times U(2)$ gauge group, are common to all models. 
 We require that the intersection  
 of these two stacks gives the three copies of the left-chiral quark doublet $Q_L$. Then, without loss of generality, 
 we may take their wrapping numbers as shown in Table 1. 
 We require that $n_2$ is coprime to 3 ($n_2 \neq 0 \bmod 3$), so that the $U(2)$ group is not replicated.
  Besides the first two stacks we have, in general,  
 three sets $I,J,K$ of $U(1)$ stacks characterised by their Chan-Paton factors: $p_i=p_2  \ \forall i \in I, \ p_1 \neq p_j \neq p_2  \ \forall j \in J$, and 
 $p_k =p_1  \ \forall k \in K$; the sets $I,J,K$ are each divided into two subsets $I_1 \cup I_2=I$ etc.,
  defined \cite{BKL1} so that the weak hypercharge $Y$ is the linear 
 combination 
 \begin{equation}
 -Y=\frac{1}{3}Q_1+\frac{1}{2}Q_2+ \sum_{i_1 \in I_1}Q_{i_1}+ \sum_{j_1 \in J_1}Q_{j_1}+ \sum_{k_1 \in K_1}Q_{k_1}
 \label{Y}
 \end{equation}
where $Q_a$ is the $U(1)$ charge associated with the stack $a$; $Q_a$ is normalised such that the ${\bf N}_a$ 
 representation of $SU(N_a)$ has $Q_a=+1$. 
  \begin{table}
\begin{center}
\begin{tabular}{|c|c|c|c|c|} \hline \hline
Stack $a$ & $N_a$ & $(n_a,m_a)$ & $(\tilde{n}_a,\tilde{m}_a)$ & $\gamma_{\theta,a}$ \\
\hline \hline
1 & 3 & $(1,0)$ & $(1,0)$ & $\alpha ^p {\bf I}_3$ \\
2 & 2 & $(n_2,3)$ & $(\tilde{n}_2,1)$ & $\alpha ^q {\bf I}_2 $\\
$i \in I$ & 1 & $(n_{i},m_{i})$ & $(\tilde{n}_i,\tilde{m}_i)$ & $\alpha ^q $\\
$j\in J$ & 1 & $(n_{j},m_{j})$ & $(\tilde{n}_j,\tilde{m}_j)$ & $\alpha ^r $\\
$k \in K$ & 1 & $(n_{k},m_{k})$ & $(\tilde{n}_k,\tilde{m}_k)$ & $\alpha ^p$ \\
\hline \hline
\end{tabular}
\end{center}
\caption{Multiplicities, wrapping numbers and Chan-Paton phases for the D5-brane models, 
($p\neq q \neq r \neq p$).}
\end{table}

 We require the cancellation of twisted tadpoles, and also that the $U(1)_Y$ gauge boson associated with the weak hypercharge 
 in eqn(\ref{Y}) does {\it not} get a mass via the generalised Green-Schwarz mechanism \cite{AFIRU1}. These requirements give the 
constraints \cite{BKL3}
\begin{eqnarray}
n_2\tilde{n}_2+ \sum _{i_1}n_{i_1}\tilde{n}_{i_1}   =   \sum _{j_1}n_{j_1}\tilde{n}_{j_1} = 1+ \sum _{k_1}n_{k_1}\tilde{n}_{k_1} \equiv T_{11} \label{T11}  \\
n_2\tilde{n}_2+ \sum _{i_2}n_{i_2}\tilde{n}_{i_2}   =  \sum _{j_2}n_{j_2}\tilde{n}_{j_2} = 2+ \sum _{k_2}n_{k_2}\tilde{n}_{k_2} \equiv T_{12} \label{T12} \\
3+ \sum _{i_1}m_{i_1}\tilde{m}_{i_1}   =   \sum _{j_1}m_{j_1}\tilde{m}_{j_1} =  \sum _{k_1}m_{k_1}\tilde{m}_{k_1} \equiv T_{41}   \label{T41}  \\
3+ \sum _{i_2}m_{i_2}\tilde{m}_{i_2}   =   \sum _{j_2}m_{j_2}\tilde{m}_{j_2} =  \sum _{k_2}m_{k_2}\tilde{m}_{k_2} \equiv T_{42}  \label{T42} \\
n_2 +\sum _{i_1}n_{i_1}\tilde{m}_{i_1}   =   \sum _{j_1}n_{j_1}\tilde{m}_{j_1} =  \sum _{k_1}n_{k_1}\tilde{m}_{k_1} \equiv T_{21}   \label{T21}  \\
n_2+ \sum _{i_2}n_{i_2}\tilde{m}_{i_2}   =   \sum _{j_2}n_{j_2}\tilde{m}_{j_2} =  \sum _{k_2}n_{k_2}\tilde{m}_{k_2} \equiv T_{22} \label{T22} \\
3\tilde{n}_2+ \sum _{i_1}m_{i_1}\tilde{n}_{i_1}   =   \sum _{j_1}m_{j_1}\tilde{n}_{j_1} =  \sum _{k_1}m_{k_1}\tilde{n}_{k_1} \equiv T_{31}   \label{T31} \\   
3\tilde{n}_2+ \sum _{i_2}m_{i_2}\tilde{n}_{i_2}   =   \sum _{j_2}m_{j_2}\tilde{n}_{j_2} =  \sum _{k_2}m_{k_2}\tilde{n}_{k_2} \equiv T_{32}  \label{T32} 
\end{eqnarray}
Open-string tachyonic scalars arise only at intersections between stacks $a$ and $b$ which have the same Chan-Paton factor $p_a=p_b$. Thus,
 Higgs doublets, which are needed to give mass to the fermionic matter, arise at $(2i_1)$ and $(2i_2)$ intersections.  We therefore 
 require also that the sets $I_1$ and $I_2$ are both non-empty. The first two constraints (\ref{T11}) and (\ref{T12}) show that $J_1 \cup K_1$
  and $J_2 \cup K_2$ are non-empty. We seek the minimal models that can satisfy all of the constraints.
  
  Suppose first that $K_1 = \emptyset$. Then eqns (\ref{T11}),(\ref{T41}),(\ref{T21}) and (\ref{T31}) show that 
  \begin{equation}
  T_{11}=1, \ T_{21}=T_{31}=T_{41}=0
  \label{Tp1j}
  \end{equation}
If there is only a single $i_1$ stack, then in general
\begin{equation}
(T_{21}-n_2)(T_{31}-3\tilde{n}_2)=n_{i_1}\tilde{m}_{i_1}m_{i_1}\tilde{n}_{i_1}=(T_{11}-n_2\tilde{n}_2)(T_{41}-3)
\end{equation}
and this is inconsistent with the values of the parameters given in eqn (\ref{Tp1j}). Thus there must be at least two stacks $i_1^{(1)},i_1^{(2)}$ 
in $I_1$. Also, since $J_1 \neq \emptyset $, we require at least one stack $j_1 \in J_1$. We label this case (1).
If, on the other hand,  $J_1 = \emptyset$, then 
 \begin{equation}
  T_{11}=T_{21}=T_{31}=T_{41}=0
  \label{Tp1k}
  \end{equation}
Thus, if there is only a single stack $i_1 \in I_1$
 \begin{eqnarray}
 I_{2i_1} \equiv n_2m_{i_1}-3n_{i_1}=T_{21}m_{i_1} =0\\
 \tilde{I}_{2i_1} \equiv \tilde{n}_2\tilde{m}_{i_1}-\tilde{n}_{i_1}=\frac{1}{3}T_{31}\tilde{m}_{i_1}=0
  \end{eqnarray}
and there are no Higgs bosons at the $(2i_1)$ intersection. So again we require at least two stacks $i_1^{(1)},i_1^{(2)} \in I_1$.
 Also, since $K_1 \neq \emptyset $, we require at least one stack $k_1 \in K_1$. This case is labelled (2).
Finally, if both $J_1 \neq \emptyset $ and $K_1 \neq \emptyset $, the minimum  content is one stack in each of $I_1,J_1$ and $K_1$.
 In \cite{BKL3} we considered models with not more than 
 one stack in each of $I_1,J_1,K_1$. There are two classes of solutions that give Higgs doublets at the $(2i_1)$ intersection:
 \begin{eqnarray}
 T_{21}=T_{41}=0, \ T_{11}=n_2p, \ T_{31}=3p \ {\rm with}  \ p = \pm 1 \\
 T_{31}=T_{41}=0, \ T_{11}=\tilde{n}_2p, \ T_{21}=p \ {\rm with}  \ p = \pm 1
 \end{eqnarray}
These are labelled (3) and (4) respectively, and in both cases we  require $n_2=-p \bmod 3$ to avoid triplication of gauge group factors.

In the same way, if $K_2 = \emptyset$, eqns (\ref{T12}),(\ref{T42}),(\ref{T22}) and (\ref{T32}) give
 \begin{equation}
  T_{12}=2, \ T_{22}=T_{32}=T_{42}=0
  \label{Tp2j}
  \end{equation}
These require at least two stacks $i_2^{(1)},i_2^{(2)} \in I_2$. Also, since $J_2 \neq \emptyset $, 
 there is at least one stack $j_2 \in J_2$. This is case (A).
On the other hand, if $J_2 = \emptyset$, then 
 \begin{equation}
  T_{12}=T_{22}=T_{32}=T_{42}=0
  \label{Tp2k}
  \end{equation}
  As before, in order to get  Higgs bosons at the $(2i_2)$ intersection, we require at least two stacks $i_2^{(1)},i_2^{(2)} \in I_2$.
 Also, since $K_2 \neq \emptyset $, we require at least one stack $k_2 \in K_2$. This is case (B). 
Finally, if both $J_2 \neq \emptyset $ and $K_2 \neq \emptyset $, the minimum is one stack in each of $I_2,J_2$ and $K_2$.
 The  two classes of models with not more than 
 one stack in each of $I_2,J_2,K_2$ and having Higgs doublets at the $(2i_2)$ intersection have \cite{BKL3}
 \begin{eqnarray}
 T_{22}=T_{42}=0, \ T_{12}=n_2q, \ T_{32}=3q \ {\rm with}  \ q = \pm 1 \\
 T_{32}=T_{42}=0, \ T_{12}=\tilde{n}_2q, \ T_{22}=q \ {\rm with} \  q = \pm 1
 \end{eqnarray}
These are labelled (C) and (D) respectively. To avoid U(2) gauge group triplication we require in (C) that $n_2=q \bmod 3$, and in (D) 
that $n_2=-q \bmod 3$. 

Evidently a minimum of eight stacks is needed to satisfy all of the constraints, and
 there are 16 possible (eight-stack) combinations of these cases obtained by combining one of cases (1),(2),(3) or (4) 
with one of (A),(B),(C) or (D). Four of these combinations, namely (3C),(3D),(4C) and (4D), which have no ``doubled'' $i_1$ or $i_2$ stacks,
 were dealt with in \cite{BKL3}. In (A) the minimal case has a single $j_2$ stack, and it is easy to see 
 from (\ref{Tp2j}) that $m_{j_2} =0= \tilde{m}_{j_2}$, and that $|n_{j_2}|$ or $|\tilde{n}_{j_2}|=2$. Thus the $U(1)$ gauge group 
 associated with this stack is unavoidably doubled, and we need consider the combinations (1A),(2A),(3A) and (4A) no further.
 
 The 8 remaining combinations may be reduced by demanding that there are no unwanted, surviving gauged $U(1)$ symmetries 
 coupled to the matter. In the first instance there are eight (real) combinations of the U(1) charges $Q_a$ coupled to the 
 twisted two-form fields $B_2^{(k)}, \ C_2^{(k)}, \ D_2^{(k)}$ and $E_2^{(k)}$ that 
 live at the orbifold singularity \cite{AFIRU1}. The two-form fields are coupled to the $U(1)_a$ field 
 strength $F_a$ of the stack $a$ by terms in the low energy 
 action of the form 
 \begin{eqnarray}
 n_a \tilde{n}_a \int _{M_4}{\rm Tr}(\gamma _{k,a} \lambda _a)B_2^{(k)} \wedge {\rm Tr}F_a  \label{nnB}\\
 m_a \tilde{m}_a\int _{M_4}{\rm Tr}(\gamma _{k,a} \lambda _a)C_2^{(k)} \wedge {\rm Tr}F_a  \\
 m_a \tilde{n}_a \int _{M_4}{\rm Tr}(\gamma _{k,a} \lambda _a)D_2^{(k)} \wedge {\rm Tr}F_a \\
 n_a \tilde{m}_a \int _{M_4}{\rm Tr}(\gamma _{k,a} \lambda _a)E_2^{(k)} \wedge {\rm Tr}F_a 
 \end{eqnarray}
 where $\gamma _{k,a} \equiv \gamma _{\theta,a}^k $ and $\lambda _a$ is the Chan-Paton matrix associated with the $U(1)$ generator. 
 For the D5-brane array given in Table 1 the coupling (\ref{nnB}) to $B_2^{(k)}$ gives a Green-Schwarz mass to the  U(1) gauge fields 
 associated with the two linearly independent charge combinations
 \begin{eqnarray}
  3Q_1+\sum_{k_1}n_{k_1}\tilde{n}_{k_1}Q_{k_1} +\sum_{k_2}n_{k_2}\tilde{n}_{k_2}Q_{k_2}-\sum_{j_1}n_{j_1}\tilde{n}_{j_1}Q_{j_1}
 -\sum_{j_2}n_{j_2}\tilde{n}_{j_2}Q_{j_2}     \\
   2n_2\tilde{n}_{2}Q_2+\sum_{i_1}n_{i_1}\tilde{n}_{i_1}Q_{i_1} +\sum_{i_2}n_{i_2}\tilde{n}_{i_2}Q_{i_2}
 -\sum_{j_1}n_{j_1}\tilde{n}_{j_1}Q_{j_1}-\sum_{j_2}n_{j_2}\tilde{n}_{j_2}Q_{j_2} 
  \end {eqnarray}
For the eight-stack models we are considering there is at most one stack in each of $J_1,K_1,J_2$ and $K_2$, so we may rewrite the 
sums over these sets in terms of the parameters $T_{11}$ and $T_{12}$. Then the two  charge combinations above become 
\begin{eqnarray}
3Q_1-Q_{k_1}-2Q_{k_2}+ T_{11}(Q_{k_1}-Q_{j_1})+ T_{12}(Q_{k_2}-Q_{j_2})  \label{B1}\\
 2n_2\tilde{n}_{2}Q_2+\sum_{i_1}n_{i_1}\tilde{n}_{i_1}Q_{i_1} +\sum_{i_2}n_{i_2}\tilde{n}_{i_2}Q_{i_2}
 -T_{11}Q_{j_1}-T_{12}Q_{j_2}
 \label{B2}
\end{eqnarray}
We may treat the other six charge combinations in the same way. In all of the models under consideration $T_{41}=0=T_{42}$,
 and in consequence only one of the two charge combinations coupled to $C_2^{(k)}$ acquires a Green-Schwarz mass, namely
 \begin{equation}
 6Q_2 +\sum_{i_1}m_{i_1}\tilde{m}_{i_1}Q_{i_1} +\sum_{i_2}m_{i_2}\tilde{m}_{i_2}Q_{i_2}
 \label{C}
 \end{equation}
 The four remaining charge combinations that acquire Green-Schwarz masses are
 \begin{eqnarray}
 T_{21}(Q_{k_1}-Q_{j_1})+T_{22}(Q_{k_2}-Q_{j_2}) \label{D1}\\
 2n_2+\sum_{i_1}n_{i_1}\tilde{m}_{i_1}Q_{i_1} +\sum_{i_2}n_{i_2}\tilde{m}_{i_2}Q_{i_2}-T_{21}Q_{j_1}-T_{22}Q_{j_2} \label{D2}\\
 T_{31}(Q_{k_1}-Q_{j_1})+T_{32}(Q_{k_2}-Q_{j_2}) \label{E1}\\
 6\tilde{n}_2+\sum_{i_1}m_{i_1}\tilde{n}_{i_1}Q_{i_1} +\sum_{i_2}m_{i_2}\tilde{n}_{i_2}Q_{i_2}-T_{31}Q_{j_1}-T_{32}Q_{j_2} \label{E2}
 \end{eqnarray}
  The model combinations (1B) and (2B), in which both $i_1$ and $i_2$ stacks are doubled, have $T_{21}=0=T_{31}=T_{22}=T_{32}$. 
  Consequently,  at most 5 of the 8 charge combinations acquire Green-Schwarz masses, leaving at least 3 remaining massless U(1) gauge bosons. 
By construction, the standard model gauge group, including $U(1)_Y$, survives as a gauge symmetry.  In addition, 
it is easy to see that the symmetry $U(1)_X$, associated with the the sum of the charges 
\begin{equation}
X=\sum_a Q_a
\end{equation}
also survives as a gauge symmetry. However, this is uncoupled to all of the matter and gauge 
fields, and so is physically unobservable. The survival of 3 or more $U(1)$ gauge symmetries entails 
the existence of at least one which is unwanted. We therefore also exclude the model combinations (1B) and (2B).
The model combinations (3B) and (4B) can be excluded on similar grounds. The fact that $T_{12}=0=T_{22}=T_{32}$ in 
 models involving (B) means that the four charge combinations (\ref{B2}),(\ref{C}),(\ref{D2}) and (\ref{E2}) are not linearly independent. 
 Further, the charge combinations (\ref{D1}) or (\ref{E1}) vanish for models (3B) or (4B) respectively.
 So again there are at least 3 massless U(1) gauge bosons. A similar argument shows that the model combinations (2C) and (2D) are also excluded. 
 Thus only the model combinations (1C) and (1D), in both of which the $i_1$ stack is doubled,  have no unwanted surviving gauged U(1) symmetry 
 at this stage. However, we must also ensure that no extra U(1) gauge symmetries arise in the way discussed in the paragraph preceding 
 eqn (\ref{Y}). This will put restrictions on the wrapping numbers for these models. 
 We return to this point later. The two models (1C) and (1D) possess the same eight stacks, namely $1,2, i_1 ^{(1)},i_1 ^{(2)}, i_2,j_1,j_2,k_2$, 
 and have $T_{11}=1,  \ T_{41}=T_{42}=T_{21}=T_{31}=0$, but differ by having $T_{22}=0$ or $T_{32}=0$ respectively.
 
 The wrapping numbers for the $i_1^{(1)}$ and $i_1^{(2)}$ stacks are obtained by solving eqns (\ref{T11}),(\ref{T41}),(\ref{T21}) and (\ref{T31})
for $\tilde{n}_{i_1}^{(1,2)}$ and $\tilde{m}_{i_1}^{(1,2)}$ in terms of the wrapping numbers  
 $n^{(1,2)}$ and $m^{(1,2)}$ on the first torus. The solutions are given in Table 2. 
Of course,  the integers $n^{(1,2)},m^{(1,2)}$ must be such that the wrapping numbers on the second torus are also integers,
 and such that there is no replication of the U(1) factors of 
either stack, as just discussed, but we do not need to write down explicit solutions with these properties at this juncture.
\begin{table}
\begin{center}
\begin{tabular}{|c|c|c|c|c|} \hline \hline
Stack $a$ &  $n_a$ & $m_a$ & $\tilde{n}_a$ & $\tilde{m}_a$  \\
\hline \hline
$i_1^{(1)} $ & $n^{(1)}$ & $m^{(1)}$ & $\delta m^{(2)} -\delta \tilde{n}_2(n_2m^{(2)}-3n^{(2)})$ & $-\delta(n_2m^{(2)}-3n^{(2)})$ \\
$i_1^{(2)} $ & $n^{(2)}$ & $m^{(2)}$ & $-\delta m^{(1)} +\delta \tilde{n}_2(n_2m^{(1)}-3n^{(1)})$ & $\delta(n_2m^{(1)}-3n^{(1)})$ \\
$j_1$  & $1$ & $0$ & $1$ & $0$ \\ 
\hline \hline
\end{tabular}
\end{center}
\caption{Wrapping numbers for the stacks $i_1^{(1)},i_1^{(2)}$ and $j_1$ in case (1). 
An arbitrary, overall 
sign $\epsilon _{j_1} = \pm 1$ is understood for the $j_1$ stack.
$\delta \equiv (n^{(1)}m^{(2)}-n^{(2)}m^{(1)})^{-1}$.}
\end{table}
Table 3 gives the wrapping numbers for cases (C) and (D) that were derived in \cite{BKL3} by requiring no 
replication of the U(1) gauge factors for the $i_2,j_2$ and $k_2$ stacks.
\begin{table}
\begin{center}
\begin{tabular}{|c|c|c|c|c|} \hline \hline
Stack $a$ &  $n_a$ & $m_a$ & $\tilde{n}_a$ & $\tilde{m}_a$  \\
\hline \hline
$i_2 $ & $n_2$ & $3$ & $q-\tilde{n}_2$ & $-1$ \\
$j_2 $ & $n_2q$ & $3q$ & $1$ & $0$  \\
$k_2 $  & $n_2q-2$ & $3q$ & $1$ & $0$ \\
\hline 
$i_2$ & $q-n_2$ & $-3$ & $\tilde{n}_2$ & $1$ \\
$j_2$ &  $1$ & $0$ & $\tilde{n}_2q $ & $q$ \\
$k_2$ &   $1$ & $0$ & $\tilde{n}_2q-2$ & $q$ \\
\hline \hline
\end{tabular}
\end{center}
\caption{Wrapping numbers for the stacks $i_2,j_2$ and $k_2$. Case (C) is at the top and case (D) at the bottom.
 A further arbitrary, overall 
sign $\epsilon _a = \pm 1$ is understood for each stack $a=i_2,j_2,k_2$.
 In both cases $q= \pm 1$. To avoid replication of the gauge groups in case (C) we require $n_2=q \bmod 3$,
  and in case (D) $n_2=-q \bmod 3$ is required.} 
\end{table}

It is straightforward to determine the spectrum. In both models there are $n_G=3$ standard model generations of chiral matter, and 
as before \cite{BKL2, BKL3}, these  include right-chiral neutrino states. Also as before, there is additional vector-like matter. 
We further constrain our models by requiring that  all of the matter, including the vector-like states, has the tree-level Yukawa couplings to 
  the tachyonic Higgs doublets at the $(2i_1^{(1,2)})$ and $(2i_2)$ intersections needed to generate mass terms via 
  the electroweak spontaneous symmetry breaking. 
 The allowed Yukawas satisfy selection rules that derive from a $Z_2$ symmetry associated with each stack of D5-branes.
 A state associated with a string 
between the $a$th and $b$th stack of D5-branes is odd under 
the $a$th and $b$th $Z_2$ and even under any other $Z_2$. 
For the  $d$ quarks the required Yukawas occur without further restriction. In both models, there are just three $d_L^c$ states, 
  and these arise at the $(1i_2)$ intersections. We have already demanded that there are Higgs doublets at the $(2i_2)$ intersections, 
  (and in both models there are three).
  Thus the required couplings of these Higgs doublets to the three $d_L^c$ states and the three quark doublets at the $(12)$ intersections {\it are} allowed.
   $u_L^c$ and possibly $\bar{u}_L^c$ states 
  arise at the $(1i_1^{(1,2)})$ intersections. Provided that there are just three $u_L^c$ states and no $\bar{u}_L^c$s, and that there are Higgs 
  doublets at the required $(2i_1^{(1,2)})$ intersections, the required Yukawas are again allowed. Thus the only constraint is that there are no 
  vectorlike $u$ quarks, and this happens provided that
  \begin{equation}
  0 \geq \bf{I}_{1i_1^{(1)}} \geq -3
  \label{nou}
  \end{equation}
 Consideration of the Yukawa couplings for the lepton states leads to stronger restrictions. Both models have six doublets $L$ at 
 the $(2k_2)$ intersections, three $\bar{L}$ doublets at the  $(2j_1)$ intersections, and three $\bar{e}_L^c$s at the $(j_1i_2)$ intersections.
  In principle, $e_L^c$ and $\bar{e}_L^c$ states 
 can arise at $(i_1^{(1,2)}j_2)$ and $(i_1^{(1,2)}k_2)$ and $(j_1i_2)$ intersections. However, any that arise at the first pair  
 will not have the required Yukawa couplings to the lepton doublets, so if we are to have mass terms for {\it all} of the matter, we 
 must ensure that there are no states at these intersections. We therefore require that
 \begin{equation}
 {\bf I}_{i_1^{(1)}j_2}=0={\bf I}_{i_1^{(2)}j_2}
 \label{ecL0}
 \end{equation}
 
 Consider first model (1C). In this case 
 \begin{equation}
 {\bf I}_{i_1^{(1)}j_2}=-q\delta (n_2m^{(1)}-3n^{(1)})(n_2m^{(2)}-3n^{(2)}) =-{\bf I}_{i_1^{(2)}j_2}
\end{equation}
There are two solutions of (\ref{ecL0}), namely
\begin{equation}
n_2m^{(1)}-3n^{(1)}=0 \  \  {\rm or} \
n_2m^{(2)}-3n^{(2)}=0  
\end{equation}
Since there is no {\it a priori} distinction between the two $i_1$ stacks, 
these two solutions will have the same physics, so 
without loss of generality we may define the stack $i_1^{(1)}$ as having wrapping numbers satisfying the first 
of these conditions. Then we see from Table 2 that $\tilde{m}_{i_1^{(2)}}=0$, so  $\tilde{n}_{i_1^{(2)}}=-\delta m^{(1)}=\eta =\pm1$ 
is required to avoid replication of the U(1) factor of the $i_1^{(2)}$ stack. It also follows that
\begin{equation}
(n^{(1)}, m^{(1)})= (n_2 \mu, 3\mu)   
\end{equation}
where $\mu = \pm 1$ to avoid replication of the U(1) factor of the $i_1^{(1)}$ stack. So $-3\delta \mu = \eta$. 
Then $\delta ^{-1} \equiv n^{(1)}m^{(2)}-n^{(2)}m^{(1)}=\mu(n_2m^{(2)}-3n^{(2)})$, and so from 
Table 2 we find that $\tilde{n}_{i_1^{(1)}}=\frac{1}{3}\eta \mu m^{(2)}-\tilde{n}_2\mu$. Thus
 $m^{(2)}=3\eta\hat{m}$ with $\hat{m}$ an integer, so that $\tilde{n}_{i_1^{(1)}}$ is an integer, and 
 $n^{(2)}=\eta(n_2\hat{m}+1)$. The wrapping numbers for the $i_1^{(1,2)}$  stacks therefore take the form given in the upper half of Table 4. 
 Similar results follow for model (1D). In this case, solving (\ref{ecL0}) requires that $m^{(1)}=0$ or $m^{(2)}=0$,
  and now we define the stack $i_1^{(1)}$ as having wrapping numbers satisfying the first 
of these conditions. Proceeding as before, we obtain the results in the lower half of Table 4.
\begin{table}
\begin{center}
\begin{tabular}{|c|c|c|c|c|} \hline \hline
Stack $a$ &  $n_a$ & $m_a$ & $\tilde{n}_a$ & $\tilde{m}_a$  \\
\hline \hline
$i_1^{(1)} $ & $n_2$ & $3$ & $-\hat{m}-\tilde{n}_2$ & $-1$ \\
$i_1^{(2)} $ & $n_2 \hat{m}+1$ & $3\hat{m}$ & $1$ & $0$ \\ 
\hline
$i_1^{(1)} $ & $1$ & $0$ & $1-n_2\tilde{n}_2+\tilde{n}_2\hat{n}$ & $-n_2+\hat{n}$ \\
$i_1^{(2)} $ & $-\hat{n}$ & $-3$ & $\tilde{n}_2$ & $1$ \\ 
\hline \hline
\end{tabular}
\end{center}
\caption{Wrapping numbers  for the stacks $i_1^{(1)},i_1^{(2)}$ satisfying eqn(\ref{ecL0}), for model (1C) at the top, 
and for model (1D) at the bottom.
 An overall, arbitrary factor of $\pm1$ for all the wrapping numbers of each stack is left undisplayed. $\hat{m}$ and $\hat{n}$ 
 are integers. }
\end{table} 

 Returning to the spectrum, using these wrapping numbers we find that for model (1C) ${\bf I}_{1i_1^{(1)}}=-3$, so the three $u_L^c$s all 
 arise at these intersections, and from eqn (\ref{nou}) there are no vector-like $u$  quarks. 
 (Recall that there are no vector-like $d$ quarks in any case.)  We explained earlier how $d$-quark mass terms arise from coupling the 
 quark doublets $Q_L$ and $d_L^c$ states to the three Higgs doublets at the $(2i_2)$ intersections. In the same way, the 
 $u$-quark mass terms arise from couplings of the quark doublets and the $u_L^c$ states to the Higgs doublets at the $(2i_1^{(1)})$ intersections.  
 The intersection numbers are
 \begin{eqnarray}
 I_{2i_1^{(1)}}=0 \ & {\rm and} & \ \tilde{I}_{2i_1^{(1)}}=\hat{m} \\
 & {\bf I}_{2i_1^{(2)}}=3 &
 \end{eqnarray}
 Then, provided $\hat{m} \neq 0$, we obtain Higgs doublets at both the $(2i_1^{(1)})$ and the $(2i_1^{(2)})$ intersections, the former 
 arising provided the distance between the parallel 1-cycles on the first torus is sufficiently small. $u$-quark  Yukawa terms coupling 
 the three $Q_L$ states at the $(12)$ intersections and the three $u_L^c$s at $(1i_1^{(1)})$ to the Higgs doublets at $(2i_1^{(1)})$ are  
 allowed by the selection rules. Similarly, for the leptons, we note first that in both models the three $\bar{L}$ 
 doublets at the $(2j_1)$ intersections and the three $\bar{e}^c_L$s at the $(j_1i_2)$ intersections may be coupled to the Higgs doublets 
 at the $(2i_2)$ intersections to give the Yukawas needed for mass terms. Using the wrapping numbers in Table 4, 
 we find for model (1C)  that ${\bf I}_{i_1^{(1)}k_2}=6$ and ${\bf I}_{i_1^{(2)}k_2}=0$, so six $e^c_L$s  arise at the first intersections 
 and none at the second. Again these couple to the six $L$ doublets at the $(2k_2)$ intersections and the $(2i_1^{(1)})$ Higgs doublets to 
 give the required Yukawa mass terms. Similarly,  there are six $\nu ^c_L$  neutrino states at the  $(i_2k_2)$ intersections and three 
 $\bar{\nu} ^c_L$s at the $(i_1^{(1)}j_1)$ intersections, which are coupled to the $L$ and $\bar{L}$ states and the Higgs doublets precisely
  as required for mass terms. The vector-like matter is therefore 
\begin{equation}
3(L+\bar{L})+3(e^c_L +\bar{e}^c_L)+3(\nu^c_L +\bar{\nu}^c_L)
\label{ng31}
\end{equation}
and all of it is coupled to the Higgs doublets as required to generate mass terms when the electroweak symmetry is spontaneously broken. 
In the case of model (1D) the  spectrum  is very similar, with the vector-like matter again given by eqn (\ref{ng31}); in this case  the three $u^c_L$s 
occur at the $(1i_1^{(2)})$ intersections, but as already noted it is a matter of convention how the two $i_1$ stacks are defined. 
Similarly, the Higgs doublet intersection numbers are
 \begin{eqnarray}
 & {\bf I}_{2i_1^{(1)}}=3 & \\
 I_{2i_1^{(2)}}=3(n_2-\hat{n}) \ & {\rm and} & \ \tilde{I}_{2i_1^{(1)}}=0 
 \end{eqnarray} 
all of which are multiples of 3. 
As already noted, there are three Higgs doublets at the $(2i_2)$ intersections in both models, and mass terms using these 
arise as in model (1C).
As before, we find that all of the matter is coupled to the Higgs doublets at these intersections
as required to generate mass terms. 

Tachyonic scalars other than the Higgs doublets are less welcome. Some of these arise at intersections
 $(ab)$ for which $I_{ab}$ or 
$\tilde{I}_{ab}$ vanishes. Such states can be removed from the low energy spectrum by making the distance between the parallel 1-cycles 
sufficiently large. The charged singlets at 
the $(i_1^{(1)}i_2)$ for model (1C), or at $(i_1^{(2)}i_2)$ for model (1D), those at the $(j_1j_2)$ intersections, 
 and the colour triplets at the $(1k_2)$ can be removed in this way.
However, the three charged-singlets at the $(i_1^{(2)}i_2)$ intersections for model (1C), or at the $(i_1^{(1)}i_2)$ intersections for model (1D),
 and the neutral singlets at 
the $(i_1^{(1)}i_1^{(2)})$ intersections cannot be removed this way, because both total intersection numbers 
${\bf I}_{i_1^{(2)}i_2}={\bf I}_{i_1^{(1)}i_1^{(2)}}=-3$ are non-zero. Nevertheless, by a judicious choice of parameters, 
we may arrange that these states are massless rather than tachyonic\footnote{
Alternatively, we might allow neutral singlets which could contribute some $e_L^c\bar{e}_L^c$ and $\nu _L^c \bar{\nu}_L^c$ mass terms,
but not $L\bar{L}$ mass terms because of the selection rules. We shall not pursue this option further.}. 
 In general, the squared mass of a tachyon at an intersection of stack $a$ with stack $b$ is given by \cite{AFIRU1}
   \begin{equation}
   m_{ab}^2=-\frac{m_{\rm string}^2}{2\pi}\left| \frac{\epsilon I_{ab}R_2/R_1}{|n_a - m_aR_2/R_1||n_b - m_bR_2/R_1|}-
   \frac{\tilde{\epsilon} \tilde{I}_{ab}\tilde{R}_2/\tilde{R}_1}{|\tilde{n}_a - \tilde{m}_a\tilde{R}_2/\tilde{R}_1||\tilde{n}_b - 
   \tilde{m}_b\tilde{R}_2/\tilde{R}_1|}\right|
   \end{equation}
where $R_1,R_2,\tilde{R}_1$ and $\tilde{R}_2$ are the radii of the fundamental 1-cycles on the two tori on which the D5-branes are wrapped;
\begin{equation}
 \epsilon \equiv 2|\cos (\theta /2)| \quad {\rm and} \quad  \tilde{\epsilon}\equiv 2|\cos (\tilde{\theta} /2)|
 \end{equation}
 with $\theta$ and $\tilde{\theta}$ the angles between the vectors defining the lattices on the two tilted tori.
 The above formula is valid provided that 
 $\epsilon$ and $\tilde{\epsilon}$ are small;  this is required in any case to ensure that the masses of the Higgs doublets are small 
 compared with the string scale. In principle, the contributions from the two tori can cancel leaving massless states rather than tachyons. 
For this to happen in model (1C) for both the charged-singlets at the $(i_1^{(2)}i_2)$ intersections and the neutral singlets at 
the $(i_1^{(1)}i_1^{(2)})$ intersections the parameters must satisfy the conditions
\begin{equation}
 \frac{\epsilon R_2/R_1}{\tilde{\epsilon}\tilde{R}_2/\tilde{R}_1}=\left| \frac{x(\hat{m}x+1)}{3(q-y)} \right|
 =\left| \frac{x(\hat{m}x+1)}{3(y+\hat{m})} \right| 
 \end{equation}
 where 
\begin{equation}
x \equiv n_2 -3R_2/R_1, \qquad y \equiv \tilde{n}_2-\tilde{R}_2/\tilde{R}_1
\end{equation}
 These are satisfied if 
 \begin{equation}
q=- \hat{m} \ {\rm or \ if} \ y= \frac{1}{2}(q- \hat{m})
 \label{mc1}
 \end{equation}
 If they {\it are} satisfied, then it is easy to see that the Higgs doublets at the $(2i_1^{(2)})$ intersections are definitely tachyonic.
Similarly, for model (1D) the cancellation occurs if 
 \begin{equation}
q= n_2-\hat{n} \ {\rm or \ if} \ x= \frac{1}{2}(q+n_2- \hat{n})
 \label{mc2}
 \end{equation}  
and again, if they are satisfied,  the Higgs doublets at the $(2i_1^{(1)})$ intersections are definitely tachyonic.

 Ratios of the gauge coupling constants are independent of the Type II string coupling constant $\lambda_{II}$. 
 For any stack $a$ with gauge coupling fine structure constant $\alpha _a$
 \begin{equation}
 \frac{\alpha_3(m_{\rm string})}{\alpha_a(m_{\rm string})}=|n_a - m_aR_2/R_1||\tilde{n}_a - \tilde{m}_a\tilde{R}_2/\tilde{R}_1|
 \end{equation}
 where $\alpha _3$ nevertheless refers to the (QCD) SU(3) gauge coupling that derives from stack $1$. 
 Thus
\begin{equation}
\frac{\alpha_3(m_{\rm string})}{\alpha_2(m_{\rm string})}=|xy|
\label{32}
\end{equation}
Also, for these models 
\begin{equation}
\frac{1}{\alpha_Y} = \frac{1}{3\alpha_3}+\frac{1}{2\alpha_2}+\frac{1}{\alpha_{i_1^{(1)}}}+\frac{1}{\alpha_{i_1^{(2)}}}+\frac{1}{\alpha_{j_1}}
\end{equation}
so for model (1C) 
\begin{equation} 
\frac{\alpha_3(m_{\rm string})}{\alpha_Y(m_{\rm string})}=\frac{1}{3}+\frac{1}{2}|xy|+|x(\hat{m}+y)| + |\hat{m}x+1| +1
\label{3Y}
\end{equation}
Consistency with a low string scale requires that these ratios do not differ greatly from the values measured \cite{cernyellow} at the 
electroweak scale $m_Z$
\begin{eqnarray}
\frac{\alpha_3(m_Z)}{\alpha_2(m_Z)} = 3.54 \\
\frac{\alpha_3(m_Z)}{\alpha_Y(m_Z)} = 11.8 \label{3Yexp}
\end{eqnarray}
It is easy to find all solutions of these:
\begin{eqnarray}
\hat{m}x =  2.08,  \ -6.62,  & 5.62, & -3.08  \\
\frac{y}{\hat{m}}  =  1.70, \  -0.53, & -0.63, & 1.15 
\label{pxy}
\end{eqnarray}
These can be satisfied with reasonable values of the parameters. For example, when $\hat{m}=1$  in the first solution we 
require $n_2 \geq 3$ to ensure that $R_2/R_1>0$. Thus, since we require that $n_2 \neq 0 \bmod3$, we
 may take $n_2=4$ and $R_2/R_1=0.64$, or 
$n_2=5$ and $R_2/R_1=0.97$. For the second choice, $q=-1=-\hat{m}$, and from eqn (\ref{mc1}) we see that 
this ensures that both the charged-singlet and neutral scalars are non-tachyonic. Similarly, for the second torus
 we may take $\tilde{n}_2=2$ and $\tilde{R}_2/\tilde{R}_1=0.3$,
 or $\tilde{n}_2=3$ and $\tilde{R}_2/\tilde{R}_1=1.3$.

For the (1D) model, eqn (\ref{3Y}) is replaced by
 \begin{equation} 
\frac{\alpha_3(m_{\rm string})}{\alpha_Y(m_{\rm string})}=\frac{1}{3}+\frac{1}{2}|xy|+|1+(\hat{n}-n_2)y|+|(x+\hat{n}-n_2)y| + 1
\label{3Y1}
\end{equation}
and the solutions are obtained by replacing $\hat{m}x$ and $y/\hat{m}$ respectively by $(\hat{n}-n_2)y$ and $x/(\hat{n}-n_2)$ in (\ref{pxy}). 
Again, it is easy to satisfy these with reasonable values of the parameters. For example, when $\hat{n}=1+n_2$ in the first solution, 
we require $\tilde{n}_2 \geq 3$ to ensure that $\tilde{R}_2/\tilde{R}_1>0$. Thus  
we may take 
$\tilde{n}_2=3$ and $\tilde{R}_2/\tilde{R}_1=0.92$,
 or $\tilde{n}_2=4$ and $\tilde{R}_2/\tilde{R}_1=1.92$, and $n_2=4$ and $R_2/R_1=0.77$, or 
$n_2=5$ and $R_2/R_1=1.1$. For the first choice of $n_2,  \ q=-1= n_2-\hat{n}$, and from eqn (\ref{mc2}) 
this ensures that both the charged-singlet and neutral scalars are non-tachyonic.

In conclusion, we find that demanding that the gauge boson  associated with weak hypercharge does  not acquire 
a string-scale mass requires intersecting D5-brane models with at least eight stacks, if we are to get the standard model gauge group 
with {\em no} additional, unwanted gauged $U(1)$ symmetries, plus three (non-supersymmetric) generations of chiral matter. This parallels 
the situation in orientifold models, in which a minimum of four stacks, plus their orientifold images, is required \cite{CIM}.
In this paper we have studied models in which we also demand that all of the matter, including vector-like leptons,
 has the Yukawa couplings to the  (open-string) 
tachyonic Higgs scalar doublets that 
are required to generate mass terms at tree level when the electroweak symmetry is spontaneously broken. 
We found two classes of model with this property. They have the same additional vector-like leptonic matter and no vector-like quark matter. 
The Higgs doublet spectrum is slightly different, in that in model (1D) the number of doublets is always a multiple of 3. Unwanted (colour-triplet, 
charged-singlet, and neutral-singlet) scalar tachyons can be removed from both models by a suitable choice of the parameters (distances between parallel 
1-cycles and wrapping numbers). These choices can be made consistently with the values needed to obtain the measured values of the ratios of the 
standard model gauge coupling constants. This is consistent with a string scale very close to the electroweak scale, which is required in  non-supersymmetric 
theories such as these in order to avoid the hierarchy problem.
The models have anomalous $U(1)$s that survive as global symmetries. Their gauge bosons acquire string-scale masses via the 
 generalised Green-Schwarz mechanism. It is expected that TeV-scale $Z'$ vector bosons will be observable at future colliders,
  and precision electroweak data (on the $\rho$-parameter) 
  already constrain \cite{GIIQ} the string scale  to be at least 1.5 TeV. 
  In particular, baryon 
  number $B=Q_1/3$ is anomalous and survives as a global symmetry. Consequently, the proton is stable despite the low string scale. 
  As before in the D4-brane case \cite{BKL1},
  the Higgs boson fields are also charged under some of the anomalous $U(1)$s that survive as global symmetries. Thus a keV-scale axion is unavoidable.

 \section*{Acknowledgements}
This research is supported in part by PPARC and the German-Israeli 
Foundation for Scientific Research (GIF).

\end{document}